\documentclass[twocolumn,prl,superscriptaddress,floatfix,showpacs,showkeys]{revtex4-1}
\usepackage[utf8]{inputenc}
\usepackage{amsmath}
\usepackage{amssymb}
\usepackage{graphicx}

\makeatletter
\usepackage{graphics}
\usepackage{epsfig}
\usepackage{color}

\newcommand{\be}{\begin{equation}} \newcommand{\ee}{\end{equation}}
\newcommand{\bea}{\begin{eqnarray}} \newcommand{\eea}{\end{eqnarray}}

\makeatother

\begin{document}

\title{Permutation entropy revisited}

\author{Stuart J. Watt} 
\author{Antonio Politi} 
\affiliation{Institute of Pure and Applied Mathematics, University of Aberdeen, Aberdeen, UK}

\date{\today}

\begin{abstract}
Time-series analysis in terms of ordinal patterns is revisited by introducing a generalized
permutation entropy $H_p(w,L)$, which depends on two different window lengths:
$w$, implicitly defining the resolution of the underlying partition;
$L$, playing the role of an embedding dimension, analogously to standard nonlinear
time-series analysis.
The $w$-dependence provides information on the structure of the corresponding invariant measure,
while the $L$-dependence helps determining the Kolmogorov-Sinai entropy.
We finally investigate the structure of the partition with the help of
principal component analysis, finding that, upon increasing $w$, the single atoms become
increasingly elongated.

\end{abstract}


\maketitle

\section{Introduction}
The development of effective procedures to encode irregular time series is an important 
research topic, tightly related to the compression of information or, equivalently, to the
identification and removal of irrelevant details within given signals.
Powerful tools have been developed when the underlying model is known and it is low-dimensional.
The state of the art is (unsurprisingly) much less satisfactory when either prior knowledge is not
available, or the dynamics is high-dimensional.
The reason can be traced back to the difficulty of explicitly partitioning the phase space into non-overlapping 
cells (atoms).

The approach proposed by Bandt and Pompe \cite{bandt02a} is the most powerful, if not the only, 
zero-knowledge method that can be effectively implemented above dimension two.
Chunks of trajectories (``windows'' as we refer to them from now on) of length $L$ are encoded 
according to the corresponding ordinal pattern (see next section for a precise definition).
The so-called permutation entropy $H_p$ is thereby determined from the probabilities of the different ordinal patterns.
In this context, a partition atom corresponds to the smallest box which contains all trajectories
encoded with the same ordinal pattern.
The easiness of the procedure has allowed developing many applications in different fields (ranging
from engineering, to medicine etc. \cite{cao04,weck15,masoller15}).

An additional reason to work with $H_p$ is its relationship with the Kolmogorov-Sinai entropy $H_{KS}$, 
the most important indicator of dynamical complexity~\cite{sinai09}.
$H_{KS}$ is a dynamical invariant, independent of the parametrization adopted to describe the underlying evolution.
$H_p$ is expected to coincide with $H_{KS}$ for sufficiently long window lengths, although the 
convergence is typically rather slow.
Recently, it has been understood that the large deviations are ``finite-size" effects
associated with the window-length dependence of the partition induced by the ordinal encoding.
These deviations can be substantially eliminated by introducing an effective permutation entropy
$\tilde H_p = H_p + D \overline{\ln \sigma}$, where $\sigma$ is the spread among trajectories characterized
by the same pattern, while $D$ is the dimension of the underlying attractor.
$\tilde H_p$ turns out to be a very accurate proxy of $H_{KS}$~\cite{politi17}.

In this paper, we revisit the concept of permutation entropy by introducing the dependence of $H_p$ on the
window length $w$ used to encode the underlying trajectory, 
while $L$ is still used to determine the entropy growth rate.
Explicit calculations of the (average) partition size confirm the intuition that the size is controlled by $w$.
This new approach allows decreasing the finite-size effects which affect the standard $H_p$, without the
need of determining the spread itself.
The spread is nevertheless investigated with the goal of characterizing the way the phase-space is filled
by the observed time series. This is done with the help of principal component analysis, by studying
the scaling  properties of the eigenvalues of the correlation matrix.

The paper is organized as follows. The general formalism is introduced in section 2.
Section 3 is devoted to the implementation of the two-length entropy, while
in Section 4, we discuss the spread of the trajectories encoded by the same symbolic sequence.
Finally in section 5, we briefly discuss possible future directions.

\section{Formalism}

Given the generic time series $(x_1,x_2,\ldots,x_n)$ (we assume it to have been properly sampled - see Ref.~\cite{kantz04}
for a discussion), a meaningful characterization requires passing through three steps:
(i) the time series must be embedded into a suitable phase space;
(ii) the corresponding space has to be properly partitioned into non overlapping cells; 
(iii) the information contained in the symbolic sequences is computed for different lengths. 

The first step is typically tackled by building an $L$-dimensional space, made of the $L$-tuples 
$(u_1,\ldots,u_L)=(x_m,x_{m+1},\ldots,x_{m+L-1})$.
Takens theorem ensures that the underlying attractor is correctly reproduced, 
provided that $L$ is large enough~\cite{takens81}.

Once the window length $L$ has been set, the next step consists in partitioning the embedding space 
into cells of size $\varepsilon$, so that the time series can be encoded as a sequence of symbols, 
each symbol corresponding to a different cell. The Kolmogorov-Sinai entropy rate $h_{KS}$ is then formally 
obtained as
\[
h_{KS} = \lim_{\varepsilon\to 0} \lim_{L\to\infty} \frac{H_{KS}(L)}{L} \; ,
\]
where the limit $\varepsilon\to 0$ is taken to ensure that the encoding is one-to-one, i.e.
to avoid that any two different, infinitely long, trajectories are encoded in the same way~\cite{eckmann85}.
If the partition is generating, this second limit is not needed. In general, there is no guarantee 
that a given partition is generating. Special approaches have been developed, which, however, work
only in two dimensions~\cite{grassberger85,christiansen97}.

In the context of permutation entropy, the $L$-tuple $(u_1,u_2,\ldots,u_L)$ is encoded as 
$S=(s_1,s_2,\ldots,s_L)$, where $s_k$ is the ordinal position of $u_k$ within the $L$-tuple.
For instance, the quadruplet $(1.3,6.1,2.5,0.7)$ is encoded as $S=(2,4,3,1)$, meaning that the 
first element is the second smallest value, and so on.
Accordingly, the phase space is automatically partitioned into cells, each containing all
$L$-tuples encoded in the same way.
The cell size $\varepsilon$ is nothing but the spread among sequences encoded in the same way; 
the spread depends on the symbolic sequence.
 
We now illustrate the process with reference to the H\'enon map, $x_{n+1}= a- x_n^2+bx_n$ 
for the standard parameter values $a=1.4$ and $b=0.3$. 
In this case, the embedding dimension $L=2$ suffices to reproduce the behavior of the 
dynamical system. We consider $L=6$ and project the partition onto a two-dimensional space. 
More precisely, given a generic 6-tuple, obtained by iterating
the H\'enon map, we plot the last two coordinates of each 6-tuple.

\begin{figure}
\centering
\includegraphics[width=0.95\linewidth,clip=true]{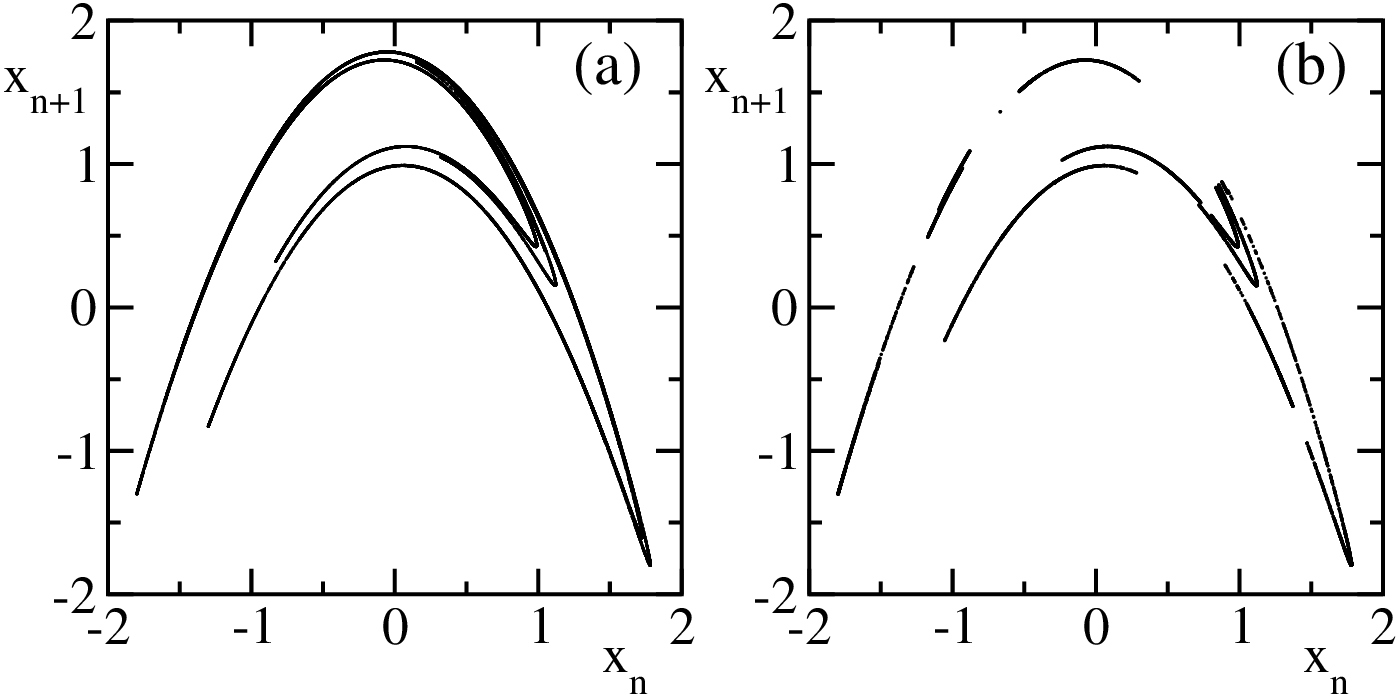}
\caption{H\'enon attractor (panel (a)); 10 stuitably chosen atoms (out of 63) of the partition
induced by the ordinal patterns of length $L=6$.}
\label{fig:partition}
\end{figure}

The results are presented in Fig.~\ref{fig:partition}. In the left panel we provide the standard representation
of the H\'enon attractor; in the right panel we plot the points belonging to 10 out of the 63 symbolic
sequences obtained by iterating the map (notice that the maximum possible number of different sequences is, in principle,
$6!$). In the picture we see a large diversity of  cell structures. In some
cases the cells are very thin and quite elongated. There is also a large diversity in the corresponding
frequencies that are only vaguely proportional to the cell size.

The beauty and, at the same time, the limit of permutation entropy is that $\varepsilon$ depends
on $L$ (actually $\varepsilon$ decreases for increasing $L$). As a result, it is sufficient to take
the limit $L\to\infty$, since it automatically implies $\varepsilon \to 0$. 
The relationship between $L$ and $\varepsilon$ is advantageous when a quick analysis is required, 
since one has to deal with only one scaling parameter.

On the other hand, the dependence of $H_p$ on $L$ induces a dependence on $\varepsilon$ as well. 
These finite-size corrections eventually vanish (in the limit $L\to\infty$), but are typically non-negligible for 
the numerically accessible $L$ values.
Moreover, the relationship between $L$ and $\varepsilon$  might represent a hindrance whenever there is no 
actual need to increase the spatial resolution, while it would instead be worth considering longer temporal 
windows.

In this paper, we revisit the definition of $H_p$, by introducing a second length, $w<L$, 
used to encode the signal; this way one can independently control the resolution $\varepsilon$.

\section{Two-length approach}

Given the $L$-tuple $(u_1,x_2,\ldots,u_L)$, we start encoding the first $w\le L$ elements 
$(u_1,x_2,\ldots,u_w)$ as in the standard implementation of permutation entropy, according to their
ordinal pattern.
Next, we proceed by encoding each following element $u_m$ up to $m=L$ according to the ordinal position
within the window $(u_{m-w+1},u_{m-w+2},\ldots,u_m)$. 
Given the pair $(w,L)$ of values, the maximum number of symbolic sequences of length $L$ is 
$w! (L-w)^w$, a number much smaller than the number $L!$ allowed by the standard approach 
(when $w\ll L$). 
This is an advantage whenever a given $w$ value provides a high-enough resolution to ensure a meaningful encoding.

Let us now denote with $p_i(w,L)$ the probability (relative frequency) of the symbolic sequence $s_i$
of length $L$, computed using an ordinal pattern of length $w$.
The corresponding {\it generalized} permutation entropy is thereby defined as,
\begin{equation}
H_p(w,L) = -\sum_{i}p_i\log p_i \; .
\end{equation} 
$H_p(L,L)$ coincides with the standard permutation entropy introduced by Pompe. 
The incremental entropy
\begin{equation}
\Delta H_p(w,L) = H_p(w,L) - H_p(w,L-1)
\label{eq:HP_gen}
\end{equation} 
is the variation of information required to characterise the time series, when the window length is increased
by one unit for a fixed partition stucture (here and in the following, we assume that the sampling time $T$ is
one unit - whenever this is not the case, one should divide the rhs by $T$).
Eq.~(\ref{eq:HP_gen}) generalizes the formula
\begin{equation}
\delta H_p(L) = H_p(L,L) - H_p(L-1,L-1)
\end{equation}
used in the context of the standard definition of permutation entropy.

\begin{figure}
\centering
\includegraphics[width=0.8\linewidth,clip=true]{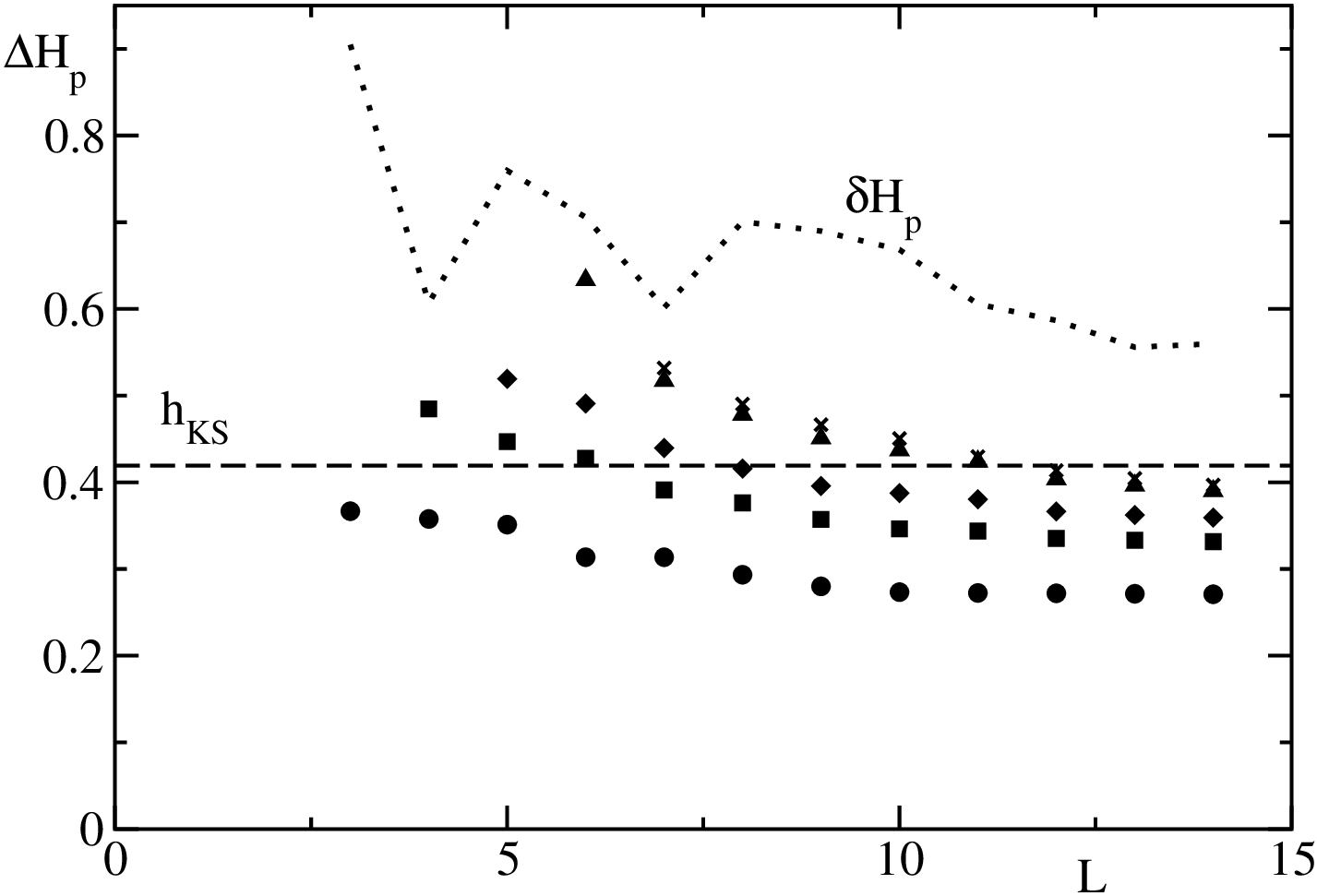}
\caption{Incremental entropy $\Delta H_p$ for the H\'enon attractor for different window lengths: circles, squares, diamonds
triangles and crosses correspond to $w=3$, 4, 5, 6, and 7, respectively. The horizontal line corresponds to the positive 
Lyapunov exponent, which coincides with the KS-entropy.}
\label{fig:inc_entropy}
\end{figure}

In Fig.~\ref{fig:inc_entropy}, we compare the two quantities with reference to the H\'enon map.
There, we see that for increasing $L$ (and $w$), $\Delta H_p$ converges faster than
$\delta H_p$ to $h_{KS}$, which coincides, in this case, with the positive Lyapunov exponent 
of the map, $\lambda_1=0.4169$.

$\Delta H_p(w,L)$ performs better than $\delta H_p$, since it corresponds to a Markov process
(of order $L-w$), while $\delta H_p$ is a hybrid observable, being the difference between two terms,
$H_p(L,L)$ and $H_p(L-1,L-1)$, which refer to different partitions and thereby to a different symbolic
encoding.

For those researchers who do not want to engage themselves in the implementation of the full two-length approach,
they can obtain a genuine and correct first-order Markov approximation by proceeding as follows.
Let $M_{ji}=p(s_j|s_i)$ denote the conditional probability to observe the sequence $s_j$ after 
shifting forward the $L$-tuple (encoded by $s_i$) by one time unit.
$M_{ji}$ can be easily estimated by determining the fraction of observed $i\to j$ transitions. 

Let us then introduce the recursive relation
\begin{equation}
q_{n+1}(s_j) = \sum_i M_{ji} q_n(s_i)
\end{equation}
where $q_n$ is a vector of probabilities (i.e. with sum-1 positive entries). 
If the underlying dynamics were a memory-1 Markov process, the numerically determined 
components $q_n$ would represent a fixed point of the above relation. 
In general, this is not the case. One can, nevertheless iterate the above equation,
(starting from a generic initial condition) until a fixed point is obtained, i.e. a vector
$q(s_j)$ that is left invariant by the above transformation.

The corresponding entropy 
\begin{equation}
K = - \sum_i q(s_i) \log q(s_i)  = \Delta H_p^{(L)}(L,L+1)
\end{equation}
coincides by construction with the first order Markov approximation 
$\Delta H_p(L-1,L) = H_p(L-1,L) - H_p(L-1,L-1)$
of the permutation entropy.

We conclude this section by discussing the dependence of $H_p(w,L)$ on $w$ for fixed $L$.
As $L$ is kept constant, it means we always refer to the same embedding dimension $L$. The variation
of the entropy is therefore due to the refinement of the partition implicitly induced by $w$.
In other words, the entropy variation is the kind of observable that is computed when a fractal
dimension is being determined within a given embedding space~\cite{kantz04}.

\begin{figure}
\centering
\includegraphics[width=0.8\linewidth,clip=true]{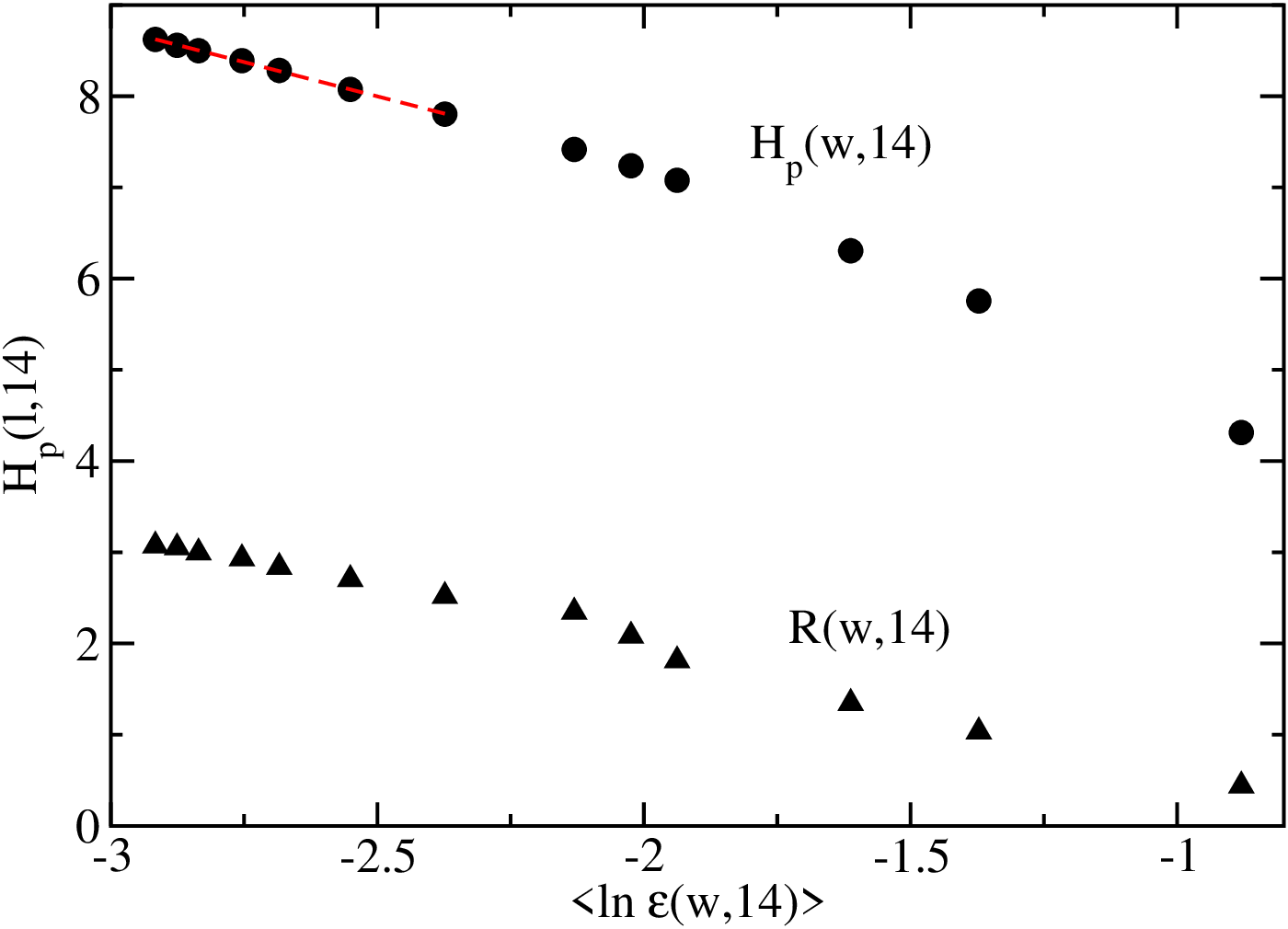}
\caption{Dependence of the entropy $H_p(w,L)$ on the cell size $\varepsilon$ for fixed $L=14$ (full circles).
From right to left, the points correspond to $w$ increasing from 2 to 14.
The ratio $R$ (see the text for its definition) is plotted for the same range of partition parameters (triangles).}
\label{fig:fractal_entropy}
\end{figure}

In order to give direct evidence of this dependence, we have estimated the spread $\varepsilon$ associated to
each symbolic sequence, by computing the standard deviation of the last variable in the corresponding $L$-tuple
(in other words, we have followed the same strategy adopted in Ref.~\cite{politi17}). 
The logarithm of the spread has been then averaged over all symbolic sequences for a given value of $w$ and $L$.
The variation of $H_P(w,L)$ with $w$ is plotted in Fig.~\ref{fig:fractal_entropy}, where, instead of referring
to $w$ itself, we treat $\langle \varepsilon \rangle(w,L)$ as the independent variable (for $L=14$). 
There, we see that the entropy increases with the logarithm of $\varepsilon$, as expected since
upon increasing $w$, the resolution used to partition a space of dimension $L$ increases as well.
A fractal structure would imply a linear growth as indeed seen in Fig.~\ref{fig:fractal_entropy}, 
where the slope (from a fit over the largest $w$-values, i.e. smallest $\varepsilon$-values) 
gives an exponent approximately equal to 1.5, 
relatively close to, but different from, the fractal dimension of the H\'enon map, $D=1.26$.

We suspect that the quantitative difference is to be attributed to the fact 
that the cells induced by the ordinal patterns are not isotropic (i.e.
characterized by a single linear size), as implicitly assumed in the definition of the fractal
dimension. We elaborate more on this point in the next section.

\section{Partition structure}
In the previous section we have shown that it is possible to improve the characterization of a complex time-series by
generalizing the encoding strategy and including the spread among equally-coded 
$L$-tuples into the analysis.

\begin{figure}
\centering
\includegraphics[width=0.8\linewidth,clip=true]{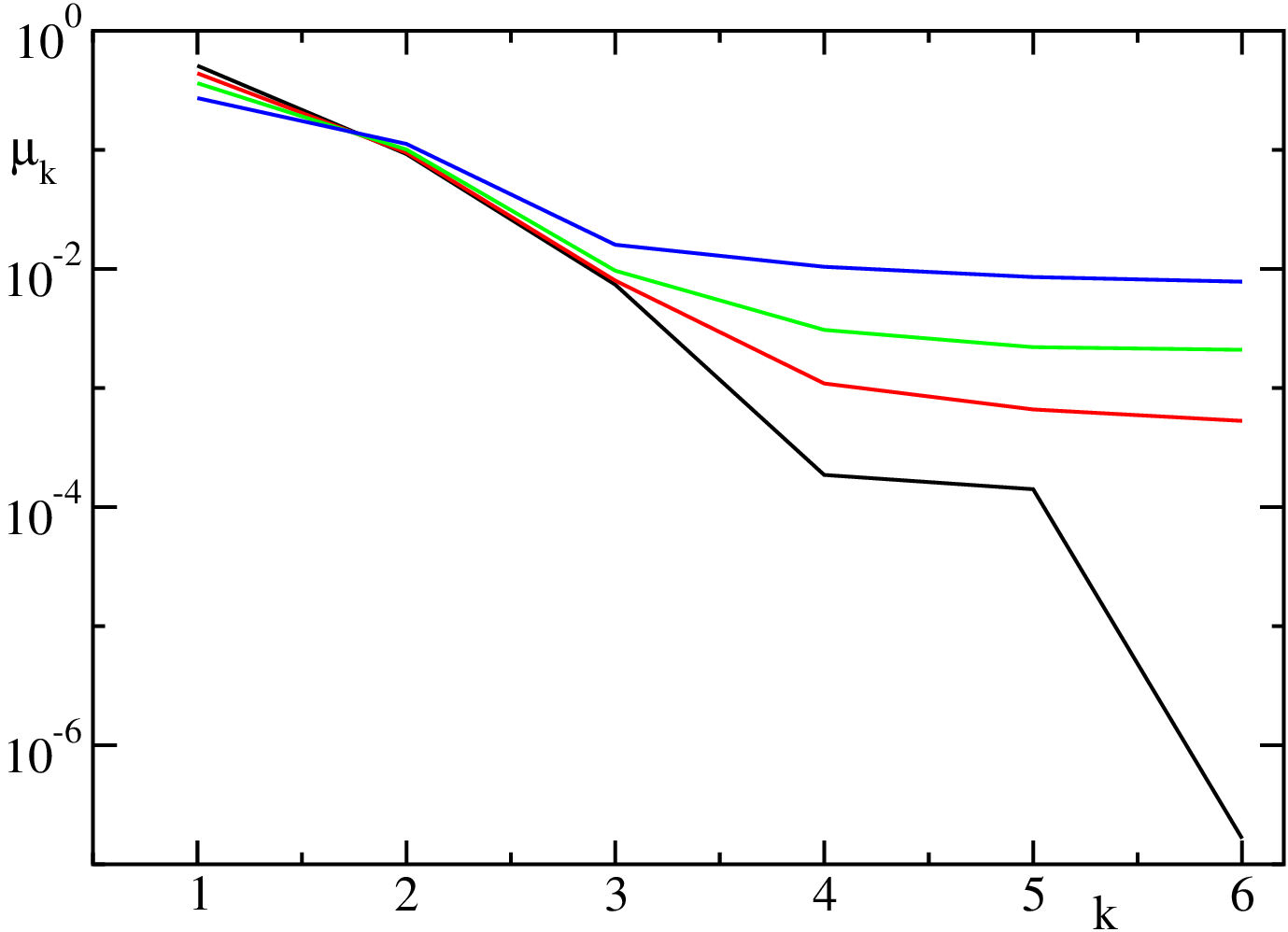}
\caption{Principal components for the H\'enon map for $L=6$. The bottom solid curve refers to the noiseless case, while
the upper curves refer to $\Delta=0.08$, $0.16$ and $0.32$, respectively.}
\label{fig:PCA_henon}
\end{figure}

In this section we analyse the distribution of points within each partition atom with the help of the
principal component analysis (PCA), alias orthogonal decomposition~\cite{broomhead86}. 
PCA is a linear tool and, as such, cannot provide an accurate
representation of an invariant measure distributed over a nonlinear manifold. Nevertheless, 
if the analysis is restricted to tiny regions, such as the atoms of the partition, the nonlinear
effects are relatively smaller and the outcome more meaningful. This approach has been already implemented
in past studies of the fractal dimension of high-dimensional systems~\cite{politi92}, with reference
to a predetermined homogeneous partition.
Here we consider the atoms induced by the ordinal representation, referring to 
the H\'enon map, for $w=L=6$.
PCA consists in first computing the covariance matrix 
$C_{ij} = \langle u_i u_j \rangle-\langle u_i\rangle\langle y_j\rangle$, where
$u_i$ denotes the $i$th component of an $L$-tuple and the average is performed
over all points lying within the same cell (i.e. encoded in the same way).
The resulting eigenvalues $\mu_k$ represent the variance of the distribution along
the so-called principal axes (the eigenvalues are assumed to be ordered from the
largest to the smallest ones).
Given such information, we further average the logarithm of $\mu_k$ for each given $k$ over all cells
(more precisely, we consider the $70\%$ most populated ones to avoid including $\mu_k$-values of
poorly populated cells).  The outcome is presented in Fig.~\ref{fig:PCA_henon}, using a logarithmic scale (see the black
solid curve at the bottom of the figure).

If one could neglect the curved nonlinear structure of the underlying attractor, only two eigenvalues should be
different from zero (due to the two-dimensional nature of the H\'enon map), while the remaining
four eigenvalues should strictly vanish.
Any deviation from zero of the third to sixth eigenvalue is therefore a manifestation of nonlinear effects 
over the scale of the cell size.
In practice we see that all six eigenvalues are different from zero although their amplitude decreases very
rapidly with the index $k$ (see the bottom solid curve).

\begin{figure}
\centering
\includegraphics[width=.8\linewidth,clip=true]{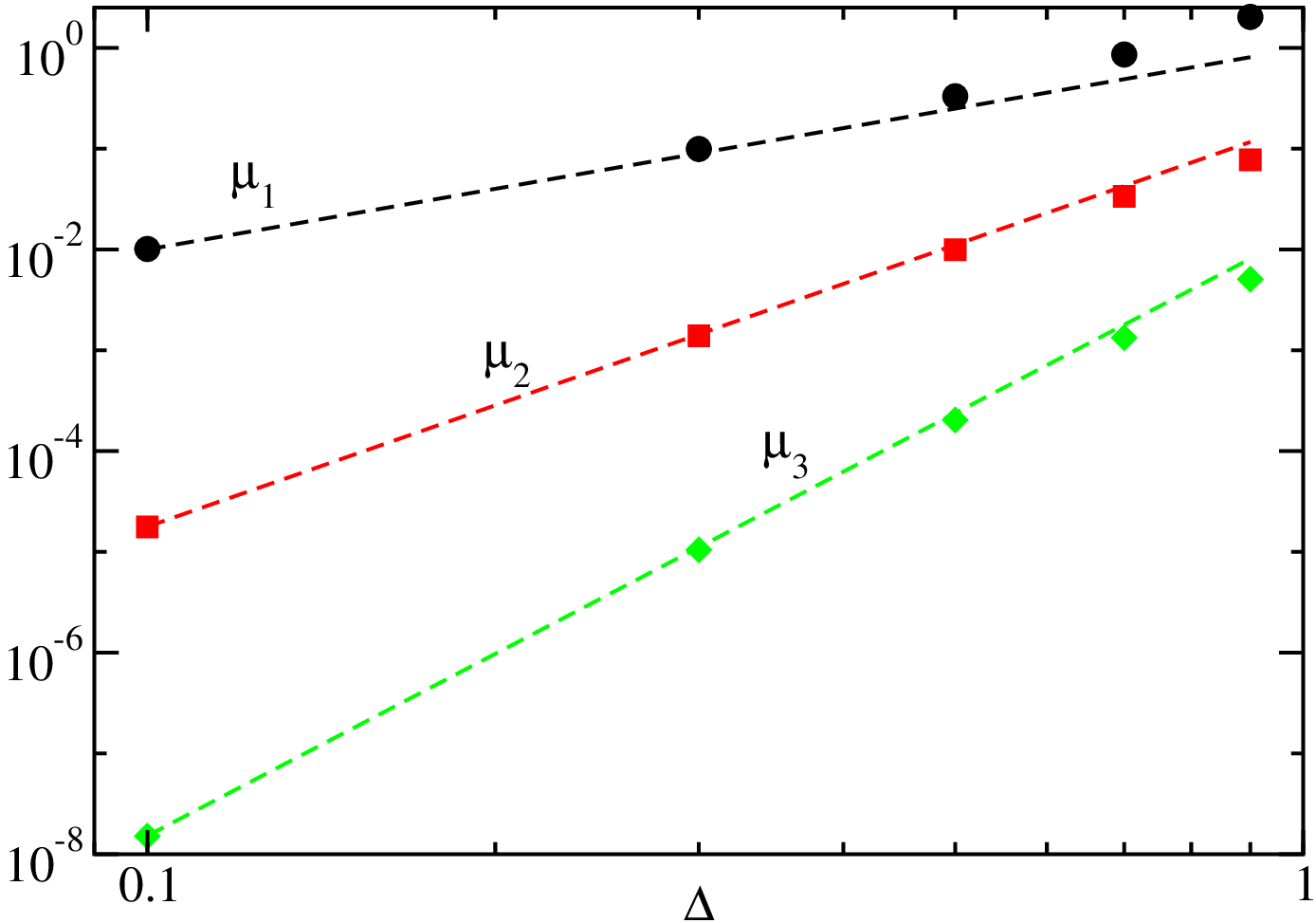}
\caption{Eigenvalues of triplets as discussed in the text for different $\Delta$ values. The dashed lines correspond
to the predicted scaling behavior.}
\label{fig:parab}
\end{figure}

In order to interpret this outcome, we turn our attention to a simple case, that can be handled analytically.
We consider a single cell in a three-dimensional space (i.e. we assume $L=3$),
filled by statistically independent triplets. Each triplet is generated by iterating twice the recursive relation 
$x_{n+1} = x_n + x_{n}^2$, starting from a randomly chosen initial condition $x_1$, 
uniformly distributed within the interval $\left[-\Delta,\Delta\right]$. Averages are then performed
over different choices of $x_1$ (rather than being time averages).
The resulting triplets are by definition aligned along a one-dimensional pseudo-parabolic curve. 
The elements of the covariance matrix $C_{ij}$ can be determined analytically by performing suitable
integrals and one can also obtain analytical expressions for the three eigenvalues. 
Rather than reporting the resulting cumbersome expressions, we
plot the $\mu$ values in Fig.~\ref{fig:parab} for different $\Delta$ values in doubly logarithmic scales 
(see full circles, crosses and triangles). Additionally, we superpose the expected scaling behavior, as
obtained from a perturbative calculation, which yields 
$\mu_1=\Delta^2$, $\mu_2= 8 \Delta^4/45$, and $\mu_3 \approx 8\Delta^6/525$ and exhibit a
very good agreement with the numerical results.

In practice, the (quadratic) nonlinearity of the initial set of points induces nonzero
eigenvalues (besides the first one). Interestingly, the higher the order $k$ of the eigenvalue,
the smaller its size. This means that the eigenvalues decrease exponentially with $k$, the
decay rate being approximately $|\ln \Delta|$ (actually, it might be even larger, because of the 
multiplicative contribution of the prefactors).
In other words, in the presence of weak nonlinearities (i.e. small $\Delta$), PCA acts as a
sort of perturbative expansion, the eigenvalues being a sort of probes which detect nonlinearities
of increasing order.

Returning back to the H\'enon map, it is resasonable to interpret the pseudo-exponential behavior 
of the eigenvalues reported in Fig.~\ref{fig:PCA_henon} as a manifestation of the nonlinear structure
of the two-dimensional manifold containing the H\'enon attractor.
Interpretative doubts, however, persist about the value of the first two eigenvalues, which both
correspond to directions actually spanned by the invariant measure. In order to partially 
clarify this point, we have computed
\begin{equation}
R^2(w,L) = \left \langle \frac{\lambda_1(w,L)}{\lambda_2(w,L)} \right \rangle \; .
\end{equation}
$R(w,L)$ is, by definition, larger than 1; it measures the degree of anisotropy of the cells
induced by the ordinal patterns. In Fig.~\ref{fig:fractal_entropy}, we plot $R(w,L)$ for
the same $w$ and $L$ values used in the computation of $H_p$ (see triangles).
Its divergence for $\varepsilon \to 0$, shows that the cells are increasingly elongated.
We suspect that this might be the origin of the overestimation of the fractal dimension.
A more quantitative analysis is however required to relate the anysotropy of the covering
with the scaling behavior of the corresponding entropy.

We finally briefly explore the role of observational noise. In Fig.~\ref{fig:PCA_henon},
we report the six eigenvalues for increasing level of noise. On the one hand, the noise has
an obvious implication: it induces a saturation of the exponential-like decrease:
the smallest eigenvalue approximately scales as $\Delta^2$, where here $\Delta$ is the noise
amplitude. On the other hand, we see a counterintuitive phenomenon:
the average leading eigenvalue decreases upon increasing $\Delta$. This effect is presumably
due to changes in the symbolic representation that more likely occur in certain
regions of the phase space than in others.

\section{Conclusions and open problems}
In this paper, we have revisited the definition of permutation entropy by generalizing the approach
proposed in Ref.~\cite{bandt02a} with the introduction of a second window-length $w$ which allows controlling the partition size.
This strategy increases the flexibility of the ordinal-pattern analysis of generic time-series; 
in particular, if combined with the measure of trajectory spreading, it allows extracting additional information 
on the structure of the invariant measure and to have hints on the presence of noise.

We have exclusively based our analysis of a prototypical example of low-dimensional chaos: the H\'enon map. 
It is certainly desirable to extend the method to higher dimensions: this is, in fact, one of the
greatest challenges in the analysis of realistic time series.
As a preliminary step in this direction, here we present results for the so-called generalized H\'enon map (GH):
$x_{n+1} = a - x_n^2 + b x_{n-2}$. This model has been already discussed in Ref.~\cite{politi17} , where it was found
to be relatively nasty (exhibiting a rather slow convergence, even compared to the higher-dimensional
attractor generated by the Mackey-Glass equation).
For $a=1.5$ and $b=0.29$, the GH map has two positive Lyapunov exponents so that the KS entropy is
equal to 0.1756 (as from the sum of the first two Lyapunov exponents).

\begin{figure}
\centering
\includegraphics[width=.8\linewidth,clip=true]{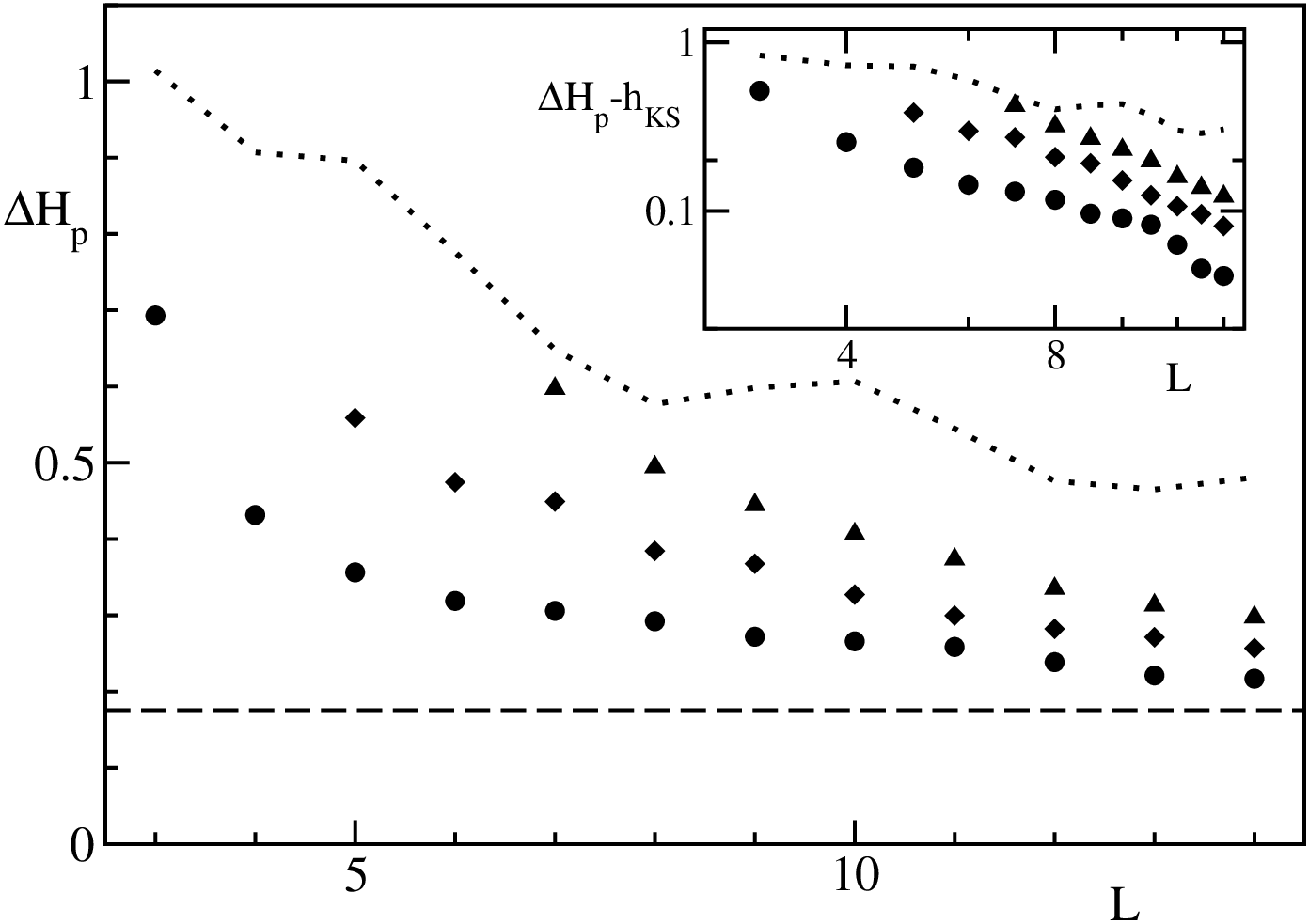}
\caption{Incremental entropy for the generalized H\'enon map ($L=14$).
Circles, diamonds and triangles correspond to $w=3$, 5, and 7, respectively.
Analogously to Fig.~\ref{fig:inc_entropy}, the horizontal dashed line corresponds
to the $h_{KS}$ as estimated from the two positive Lyapunov exponents, while the
dotted line corresponds to the standard implementation of the permutation entropy.
In the inset, the difference $\Delta H_p-\delta h_{KS}$ is ploted in doubly
logarithmic scales to appreciate the convergence rate.}
\label{fig:genhen}
\end{figure}

The results presented in Fig.~\ref{fig:genhen} confirm that the two-length approach is superior
to the computation of the standard permutation entropy. However, the convergence to the asymptotic
value is slower and, more important, it seems to follow a weird pattern. In fact, smaller $w$ values
seem to yield better results: compare, for instance, full circles ($w=3$) to triangles ($w=7$).
As it can be seen from the inset, where the deviation from the asymptotic value is plotted versus
$L$ in doubly logarithmic scales, all sets of measurement are compatible with the final value.
The reason of the lower performance of the supposedly more accurate partitions need to be further
clarified. Anyway, this ``anomalous" scenario is consistent with the slowness of the convergence
reported in Ref.~\cite{politi17}.

Altogether, the method proposed in this paper is significantly more accurate than the standard
one, but there are many issues that require additional investigations: what is the reason for the
slow convergence exhibited by the GH map? Is it a peculiarity of the model itself, or a
general feature of some broad class of high-dimensional dynamics?
Moreover, can we quantify the effect of noise so as to distinguish genuine deterministic
from stochastic contributions?

\acknowledgments
One of us, (SJW), wishes to acknowledge financial support from the Carnegie Trust for his summer
project.


\end{document}